\documentstyle[12pt,epsf]{article}
\begin{document}
\large

\begin{center}
{\bf Neutrino Oscillations. Theory and Experiment}\\
\vspace{1cm}
\par
Beshtoev Kh. M.\\
\vspace{1cm}
\par
Joint Institute for Nuclear Research, Joliot Curie 6, 141980 Dubna,
Moscow region, Russia\\

\end{center}

\par
{\bf Abstract}\\

\par
The theoretical schemes on neutrino oscillations are considered. The
experimental data on neutrino oscillations from Super-Kamiokande (Japan)
and SNO (Kanada) are given. Comparison of these data with theoretical schemes
is done. Conclusion is made that the experimental data have confirmed the
scheme only with transitions (oscillations) between aromatic
$\nu_e, \nu_\mu, \nu_\tau$ neutrinos with maximal mixing angles.\\



\section{Introduction}
\par
The suggestion that, by analogy with $K^{o},\bar K^{o}$ oscillations,
there could be
neutrino oscillations (i.e., that there could be neutrino-antineutrino
oscillations $\nu \rightarrow \bar \nu$) was considered by Pontecorvo [1]
in 1957. It was subsequently considered by Maki et al. [2] and Pontecorvo [3]
that there could be mixings (and oscillation) of neutrinos of different aromas
(i.e., $\nu _{e} \rightarrow \nu _{\mu }$ transitions).
\par
The problem of solar neutrinos arose after the first experiment performed
to measure the flux of neutrinos from the Sun by the $^{37}Cl - ^{37}Ar$
[4] method. The flux was found to be several times smaller than expected
from calculations made in accordance with the standard solar model (SSM) [5].
It was suggested in [6] that the solar neutrino deficit could be explained by
neutrino oscillations.
Subsequently, when the result of the experiment at Kamiokande [7] confirmed
the existence of the deficit relative to the SSM calculations, one of the
attractive approaches to the explanation of the solar neutrino deficit became
resonant enhancement of neutrino oscillations in matter [8].
Resonant enhancement of neutrino oscillations in matter was obtained from
Wolfenstein's equation for neutrinos in matter [9]. It was noted in Ref. [10]
that Wolfenstein's equation for neutrinos in matter is an equation for
neutrinos in matter in which they interact with matter not through the weak
but through a hypothetical weak interaction that is
left-right symmetric. Since in the standard weak interactions participate
only left components of neutrinos the results obtained from Wolfenstein's
equation have no direct relation to real neutrinos.
\par
Later experimentalists obtained the first results on
the Gran Sasso $^{71}Ga - ^{71}Ge$  experiment [11], that within a
$3\sigma $ limit did not disagree with the SSM calculations.
The new data from the SAGE experiment [12] are fairly close to the Gran Sasso
results.
\par
In Ref. [13], the author of this article proposed a new mechanism of
enhancement of
neutrino oscillations in matter that is realized through the weak interaction
of oscillation neutrinos with matter if the thickness of this matter is
sufficiently great. Later in works [14] it was shown that since the
standard weak interactions  cannot generate masses, the resonance enhancement
of neutrino oscillations in matter cannot be realized without violation of
the energy-momentum conservation law.
\par
Besides the experimental devices marked above, at present there are working
Super-Kamiokande [15-17] and SNO [18] detectors. The experimental results
obtained with SNO detector present a great interest since they can be used for
modelness analysis of neutrino oscillations.
\par
After the discovery of neutrino oscillations on Super-Kamiokande [19] (by non
direct method) and on SNO [20] (by direct method) it is necessary to
analyze the situation which arises in the problem of neutrino oscillations.
\par
In this work theoretical schemes of neutrino oscillations and their analyses
are considered. Also the experimental data obtained on Super-Kamiokande (Japan)
and SNO (Canada) are given. Comparison of these data with consequences
in the theoretical schemes has been carried out.

\section{Theory}

\par
{\bf 2.1  Distinguishing Features of Weak Interactions}\\

\par
The strong and electromagnetic interaction theories are left-right systemic
theories (i.e. all components of the spinors participate in these interactions
symmetrically). In contrast to this
only the left components of fermions participate
in the weak interaction. We will consider some consequences deduced from this
specific feature of the weak interaction.
\par
The local conserving current $j^{\mu i}$
of the weak interaction has the following form:
$$
j^{\mu i} = \bar \Psi_L \tau^i \gamma^\mu \Psi_L  ,
\eqno(1)
$$
where $\bar \Psi_L, \Psi_L$ are lepton or quark doublets
$$
\left(\begin{array}{c} e\\ \nu_e \end{array}\right)_{i L}
\eqno(2)
$$
$$
\left(\begin{array}{c} q_1\\ q_2 \end{array}\right)_{i L},\qquad i = 1-3  ,
$$
where $i$ is aromatic number of quarks or leptons.
\par
The currents $S^\mu_i$ obtained from  the global abelian transformation
by using Neuter theorem [21] are
$$
S^\mu_i = i (\bar \Psi_i \partial_\mu \Psi_i) ,
\eqno(3)
$$
(where $i$ characterizes the type of the gauge  transformation) and
the corresponding
conserving current (the forth component of $S^\mu_i$) is
$$
I_i =  \int S^0_i d^3x = \int \epsilon \bar \Psi_i \Psi_i d^3x  ,
\eqno(4)
$$
where $\epsilon$ is the energy of fermion $\Psi_i$.
\par
Since we cannot switch off the weak interactions, then while the particle
is moving in vacuum all the effects connected with these interactions will be
realized.
\par
If now we take into account that the right components
of fermions $\bar \Psi_{i R}, \Psi_{i R}$ do not participate in the
weak interaction, then from (4) for abelian currents we get
$$
I_i = \int \epsilon \bar \Psi_{i L} \Psi_{i L} d^3 x  \equiv 0  ,
\eqno(5)
$$
i.e. (in contrast to the strong and electromagnetic
interactions) no conserving additive numbers appear in the
weak interaction. However, we can see from experiments that
the hierarchical violation of these additive numbers takes place here
(see [22] and references there).\\

\newpage
\par
{\bf 2.2 About Neutrino mass}\\

\par
a) Hypothesis: A massless free particle cannot have a charge.
An example of this case is photon (carrier of the electromagnetic interactions),
which has no charge. To gluons, which are in the confining state, this hypothesis
cannot be applied. In application to neutrino having a weak charge, this
hypothesis drives to a conclusion: the neutrino participating in weak
interactions cannot be massless. In work [23] this hypothesis at sufficiently
common suppositions was proved.
\par
b) The discovery of neutrino oscillations is an additional confirmation of the
conclusion that it is a massive particle.\\

\par
{\bf 2.3 Theory of Neutrino Oscillations}\\

\par
In the old theory of
neutrino oscillations [24, 6], constructed in the framework of Quantum
theory in analogy with the theory of $K^{o}, \bar{K}^{o}$ oscillation,
it is supposed that mass eigenstates are $\nu_{1}, \nu_{2}, \nu_{3}$
neutrino states but not physical neutrino states
$\nu_{e}, \nu_{\mu }, \nu_{\tau}$, and that the neutrinos
$\nu_{e}, \nu_{\mu }, \nu_{\tau}$ are created as superpositions
of $\nu_{1}, \nu_{2}, \nu_{3}$  states.  This means that the
$\nu_{e}, \nu_{\mu }, \nu_{\tau}$ neutrinos have no definite mass, i.e.
their masses may vary in dependence on the  $\nu_{1}, \nu_{2}, \nu_{3}$
admixture  in the $\nu_{e}, \nu_{\mu }, \nu_{\tau}$  states.
Naturally, in this case the law of conservation of the energy and
the momentum of the neutrinos is not fulfilled. Besides, every particle
must be created on its mass shell and it will be left on its mass shell
while passing through vacuum. It is clear that this picture is incorrect.
\par
In the modern theory on neutrino oscillations [25]-[26], constructed in
the framework of the particle physics theory it is supposed that:
\par
1)  The  physical  states  of  the $\nu_{e}, \nu_{\mu }, \nu_{\tau}$
neutrinos   are eigenstates of the weak interaction and, naturally,
the mass
matrix of $\nu_{e}, \nu_{\mu }, \nu_{\tau}$ neutrinos is diagonal.
All  the  available, experimental results indicate  that  the  lepton
numbers $l_{e}, l_{\mu }, l_{\tau}$  are   well conserved, i.e. the standard
weak interactions do  not  violate  the lepton numbers.
\par
2) Then, to violate the  lepton  numbers, it  is  necessary  to introduce an
interaction violating these numbers. It is  equivalent to introducing
nondiagonal  mass terms  in the  mass  matrix  of $\nu_{e}, \nu_{\mu },
\nu_{\tau}$. Diagonalizing this matrix we go to the $\nu_{1}, \nu _{2},
\nu_{3}$ neutrino states. Exactly like the case  of $K^{o}$  mesons
created  in strong interactions, when mainly $K^{o}, \bar{K}^{o}$
mesons are produced, in  the considered case $\nu_{e}, \nu_{\mu }, \nu_{\tau}$,
but not $\nu_{1}, \nu_{2}, \nu_{3}$, neutrino  states  are mainly created
in the weak interactions (this is so, because   the contribution of the
lepton numbers violating interactions  in this process is too small).
And in this case no oscillations take place.
\par
3) Then, when the $\nu_{e}, \nu_{\mu }, \nu_{\tau}$  neutrinos are pass
through vacuum, they  will  be  converted  into superpositions  of  the
$\nu_{1}, \nu _{2}, \nu_{3}$  owing  to the presence  of  the
interactions violating  the  lepton number of neutrinos and  will be left
on  their mass   shells.  And, then, oscillations of the
$\nu_{e}, \nu_{\mu}, \nu_{\tau}$ neutrinos will  take  place according to
the standard scheme [24-26]. Whether these oscillations are real
or virtual, it will be determined by the masses of the  physical neutrinos
$\nu_{e}, \nu_{\mu}, \nu_{\tau}$.
\par
i) If the masses of the $\nu_{e}, \nu_{\mu }, \nu_{\tau}$ neutrinos  are
equal, then the real oscillation of the neutrinos will take  place.
\par
ii) If  the masses  of  the $\nu_{e}, \nu _{\mu }, \nu _{\tau}$ are  not
equal, then the virtual oscillation of  the  neutrinos will  take place.
To make these oscillations  real,  these  neutrinos must participate  in the
quasielastic interactions, in order to undergo transition  to  the mass shell
of the other appropriate neutrinos by analogy with $\gamma  - \rho ^{o}$
transition  in the  vector   meson  dominance model.  In case  ii)
enhancement of neutrino oscillations will  take place  if the mixing angle is
small at neutrinos passing through  a bulk of matter [13, 27].
\par
So, the mixings (oscillations) appear since
at neutrinos creating
eigenstates of the weak interaction are realized
(i.e. $\nu_e, \nu_\mu, \nu_\tau$ neutrinos)
but not the eigenstates of the weak interaction violating
lepton numbers (i.e. $\nu_1, \nu_2, \nu_3$ neutrinos) and then,
when passing through vacuum, they are converted into superpositions of
$\nu_1, \nu_2, \nu_3$ neutrinos.
If $\nu_1, \nu_2, \nu_3$ neutrinos were originally created, then the mixings
(oscillations) would not have taken place since the weak interaction conserves
the lepton numbers.
\par
Now we come to a more detailed consideration of the oscillations.
For simplification we consider the oscillation of two types of neutrinos
$\nu_e, \nu_\mu$ having $l_{\nu_e}, l_{\nu_\mu}$ numbers
which can transit each into the other.
We can use the mass matrix of $\nu_e, \nu_\mu$ neutrinos to consider
transitions between these particles in the framework of the quantum
theory (or particle physics) since the mass matrix is an eigenstate of the type
of interaction which creates these particles (see below).
\par
The mass matrix of $\nu_e$ and $\nu_\mu$ neutrinos has the form
$$
\left(\begin{array}{cc} m_{\nu_e}& 0 \\ 0 & m_{\nu_\mu}  \end{array}
\right) .
\eqno(6)
$$
\par
Due to the presence of the interaction
violating the lepton numbers, a nondiagonal term appears in this
matrix and then this mass matrix is transformed into the following
nondiagonal matrix ($CP$ is conserved):
$$
\left(\begin{array}{cc}m_{\nu_e}
& m_{\nu_e \nu_\mu} \\ m_{\nu_\mu \nu_e} & m_{\nu_\mu} \end{array}
\right) ,
\eqno(7)
$$
then the lagrangian of mass of the neutrinos takes the following form
($\nu \equiv \nu_L$):
$$
\begin{array}{c}{\cal L}_{M} = - \frac{1}{2} \left[m_{\nu_e}
\bar \nu_e \nu_e
+ m_{\nu_\mu} \bar \nu_{\mu} \nu_{\mu }
+ m_{\nu_e \nu_{\mu }}(\bar \nu_e \nu_{\mu } + \bar \nu_{\mu }
\nu _e) \right] \equiv \\
\equiv  - \frac{1}{2} (\bar \nu_e, \bar \nu_\mu)
\left(\begin{array}{cc} m_{\nu_e} & m_{\nu_e \nu_{\mu }} \\
m_{\nu_{\mu} \nu_e} & m_{\nu_\mu} \end{array} \right)
\left(\begin{array}{c} \nu_e \\ \nu_{\mu } \end{array} \right)
\end{array} ,
\eqno(8)
$$
which is diagonalized by turning through  the
angle $\theta$ and (see ref. in [24]) and then this lagrangian (8) transforms
into the following one:
$$
{\cal L}_{M} = - \frac{1}{2}
\left[ m_{1} \bar \nu_{1} \nu_{1} + m_{2} \bar \nu_{2} \nu_{2} \right]  ,
\eqno(9)
$$
where
$$
m_{1, 2} = {1\over 2} \left[ (m_{\nu_e} + m_{\nu_\mu}) \pm
\left((m_{\nu_e} - m_{\nu_\mu})^2 +
4 m^{2}_{\nu_\mu \nu_e} \right)^{1/2} \right] ,
$$
\par
\noindent
and angle $\theta $ is determined by the following expression:
$$
tg 2 \theta  = \frac{2 m_{\nu_e \nu_\mu}} {(m_{\nu_\mu}
- m_{\nu_e})} ,
\eqno(10)
$$
$$
\begin{array}{c}
\nu_e = cos \theta  \nu_{1} + sin \theta \nu_{2}  ,         \\
\nu _{\mu } = - sin \theta  \nu_{1} + cos \theta  \nu_{2} .
\end{array}
\eqno(11)
$$
From eq.(10) one can see that if $m_{\nu_e} = m_{\nu_{\mu}}$,
then the mixing angle is equal to $\pi /4$ independently of the value of
$m_{\nu_e \nu_\mu}$:
$$
sin^2 2\theta = \frac{(2m_{\nu_{e} \nu_{\mu}})^2}
{(m_{\nu_e} - m_{\nu_\mu})^2 +(2m_{\nu_e \nu_{\mu}})^2} ,
\eqno(12)
$$
$$
\left(\begin{array}{cc} m_{\nu_1} & 0 \\
0 & m_{\nu_2} \end{array} \right) .
$$

\par
It is interesting to remark that expression (12) can be obtained from
the Breit-Wigner distribution [28]
$$
P \sim \frac{(\Gamma/2)^2}{(E - E_0)^2 + (\Gamma/2)^2}   ,
\eqno(13)
$$
by using the following substitutions:
$$
E = m_{\nu_e},\hspace{0.2cm} E_0 = m_{\nu_\mu},\hspace{0.2cm}
\Gamma/2 = 2m_{\nu_e, \nu_\mu} ,
$$
where $\Gamma/2 \equiv W(... )$ is a
width of $\nu_e \rightarrow \nu_\mu$ transition, then we can use a standard
method [26, 29] for computing this value.
\par
The expression for time evolution of $\nu _{1}, \nu _{2}$ neutrinos
(see (9), (11)) with masses $m_{1}$ and $m_{2}$ is
\par
$$
\nu _{1}(t) = e^{-i E_1 t} \nu _{1}(0),  \qquad
\nu _{2}(t) = e^{-i E_2 t} \nu _{2}(0) ,
\eqno(14)
$$
where
$$
E^2_{k} = (p^{2} + m^2_{k}), k = 1, 2 .
$$
\par
If neutrinos are propagating without interactions, then
\par
$$
\begin{array}{c}
\nu_e(t) = cos \theta e^{-i E_1 t} \nu_{1}(0) + sin \theta
e^{-i E_2 t} \nu_{2}(0) , \\
\nu_{\mu }(t) = - sin \theta e^{-i E_1 t} \nu_{1}(0) + cos \theta
e^{-i E_2 t} \nu_{2}(0) .
\end{array}
\eqno(15)
$$
\noindent
Using the expression for $\nu _{1}$ and $\nu _{2}$  from  (11),
and putting it into (15), one can get the following expression:
$$
\nu_e (t) = \left[e^{-i E_1 t} cos^{2} \theta + e^{-i E_2 t} sin^{2} \theta
\right] \nu _e (0) +
$$
$$
+ \left[e^{-i E_1 t} - e^{-i E_2 t} \right] sin \theta \cos \theta
\nu_{\mu }(0) ,
\eqno(16)
$$
$$
\nu_{\mu }(t) = \left[e^{-i E_1 t} sin^{2} \theta + e^{-i E_2 t}
cos^{2} \theta \right] \nu_{\mu}(0)  +
$$
$$
+ \left[e^{-i E_1 t} - e^{-i E_2 t} \right] sin\theta cos \theta \nu_e (0) .
$$
\par
The probability that neutrino $\nu_e$ created at the time $t = 0$ will be
transformed into $\nu_{\mu}$ at the time $t$ is an absolute value of
amplitude $\nu_{\mu}(0)$ in (16) squared, i. e.
\par
$$
\begin{array}{c}
P(\nu_e \rightarrow \nu_{\mu}) = \mid(\nu_{\mu}(0) \cdot \nu_e(t)) \mid^2 =\\
 = {1\over 2} \sin^{2} 2\theta \left[1 - cos ((m^{2}_{2} - m^{2}_{1}) / 2p)
t \right] ,
\end{array}
\eqno(17)
$$
\noindent
where it is supposed that $p \gg  m_{1}, m_{2}; E_{k} \simeq
p + m^{2}_{k} / 2p$.
\par
The expression (17) presents the probability of neutrino aroma oscillations.
The angle $\theta$ (mixing angle) characterizes value of mixing.
The probability $P(\nu_e \rightarrow  \nu_{\mu})$ is a periodical function
of distances where the period is determined by the following expression:
$$
L_{o} = 2\pi  {2p \over {\mid m^{2}_{2} - m^{2}_{1} \mid}} .
\eqno(18)
$$
\par
And probability $P(\nu _e \rightarrow  \nu _e)$ that the neutrino $\nu_e$
created at time $t = 0$ is preserved as $\nu_e$ neutrino at time $t$ is
given by the absolute value of the amplitude of $\nu_e(0)$  in (16) squared.
Since the states in (16) are normalized states, then
$$
P(\nu_e \rightarrow  \nu_e) + P(\nu_e \rightarrow \nu_{\mu}) = 1 .
\eqno(19)
$$
\par
So, we see that aromatic oscillations caused by nondiagonality of the neutrinos
mass matrix violate the law of the $-\ell_e$ and $\ell_{\mu}$ lepton number
conservations. However in this case, as one can see from exp. (19), the full
lepton numbers $\ell  = \ell_e + \ell_{\mu}$ are conserved.
\par
We can also see that there are two cases of $\nu_e, \nu_\mu$
transitions (oscillations) [26], [29].
\par
1. If we consider the transition of $\nu_e$ into
$\nu_\mu$ particle, then
$$
sin^2 2\beta \cong \frac{4m^2_{\nu_e, \nu_\mu}}{(m_{\nu_e} - m_{\nu_\mu})^2
+ 4m^2_{\nu_e, \nu_\mu}}  ,
\eqno(20)
$$
if the probability of the transition
of $\nu_e$ particles into $\nu_\mu$ particles through the interaction
(i.e. $m_{\nu_e, \nu_\mu}$) is very small, then
$$
sin^2 2\beta \cong \frac{4m^2_{\nu_e, \nu_\mu}}{(m_{\nu_e} -
m_{\nu_\mu})^2} \cong 0 .
\eqno(21)
$$
\par
How can we understand this  $\nu_e \rightarrow \nu_\mu$ transition?
\par
If $2m_{\nu_e, \nu_\mu} = \frac{\Gamma}{2}$ is not zero, then it means
that the mean mass of $\nu_e$ particle is $m_{\nu_e}$ and
this mass is distributed by $sin^2 2\beta$ (or by the Breit-Wigner
formula) and the probability of the $\nu_e \rightarrow \nu_\mu$ transition
differs from zero and it is defined by masses of $\nu_e$ and $\nu_\mu$
particles and $m_{\nu_e, \nu_\mu}$, which is computed in the framework
of the standard method, as pointed out above.
\par
So, this is a solution of the
problem of the origin of mixing angle in the theory of vacuum oscillations.
\par
In this case the probability of $\nu_e \rightarrow \nu_\mu$ transition
(oscillation) is described by the following expression:
$$
P(\nu_e \rightarrow \nu_\mu, t) =  sin^2 2\beta sin^2
\left[\pi t\frac{\mid m_{\nu_1}^2 - m_{\nu_2}^2 \mid}{2 p_{\nu_e}} \right ] ,
\eqno(22)
$$
where $p_{\nu_e}$ is a momentum of $\nu_e$ particle.
\par
Originally it was supposed [6, 24] that these oscillations are real oscillations.
However we see that these oscillations are virtual because  when $\nu_e$
really transits into $\nu_\mu$, then it can decay into electron neutrino plus
something, i.e. we gain the energy from vacuum, which is equal to
the mass difference $\Delta m = m_{\nu_\mu} - m_{\nu_e}$ (momenta of
$\nu_e$ and $\nu_\mu$ are equal at oscillations). Then it is clear that
at real $\nu_e \to \nu_\mu$
transition the law of energy conservation is violated. This law can be
fulfilled only at virtual $\nu_e \to \nu_\mu$ transitions.
\par
2. If we consider the virtual transition of $\nu_e$ into $\nu_\mu$ neutrino
at  $m_{\nu_e} = m_{\nu_\mu}$ (i.e. without changing the mass shell), then
$$
tg 2\beta = \infty  ,
$$
$\beta = \pi/4$, and
$$
sin^2 2\beta = 1     .
\eqno(23)
$$
\par
In this case the probability of the $\nu_e \rightarrow \nu_\mu$ transition (oscillation) is
described by the following expression:
$$
P(\nu_e \rightarrow \nu_\mu, t) =
\left[\pi t\frac{4 m_{\nu_e, \nu_\mu}^2}{2 p_a} \right ] .
\eqno(24)
$$
\par
To make these virtual oscillations real, their participation in quasielastic
interactions is necessary for the transitions to their own mass shells [29].
\par
It is clear that the $\nu_e \rightarrow \nu_\mu$ transition is
a dynamical process.
\par
Now let us consider the common case. In this case the mass lagrangian
has the following form:
\par
$$
\begin{array}{c}
{\cal L}_{M} = - \bar \nu_{R} M \nu_{L} + H. c. \equiv \\
\equiv \sum^{}_{l,l'= e, \mu ,\tau} \nu_{l'R} M_{l'l} \nu_{lL}
+ H. c.   ,
\end{array}
\eqno(25)
$$
\par
\noindent
$M$ is a complex $3\times3$ matrix.  It is necessary to remark that the $\nu_{R}$
is absent in the weak interactions lagrangian. By using the expression
$$
M = V m U^{+} ,
\eqno(26)
$$
\par
\noindent
(where $V, U$ - unitary matrices) we transform ${\cal L}_{M}$ to a diagonal
form
\par
$$
\begin{array}{c}
{\cal L}_M  = - \bar \nu_{R} m \nu_{L} + H. c. \equiv \\
\equiv \sum^{3}_{k=1} m_{k} \bar \nu_{k} \nu_{k} + H. c. ,
\end{array}
\eqno(27)
$$
\par
\noindent
where
$$
m_{i k} = m_{k} \delta_{i k} ,
$$
and
$$
\nu^{'}_{L} = U^{+}\nu_{L},\quad \nu^{'}_{R} = V^{+} \nu _{R},\quad
\nu^{'} = \left(\begin{array}{c} \nu_1 \\ \nu_2 \\ \nu_3 \end{array} \right).
\eqno(28)
$$
\noindent
We can see that the lagrangian (25) is invariant at the global gauge
transformation
$$
\nu_{k}(x) \rightarrow  e^{\Lambda} \nu _{k}(x)
\eqno(29)
$$
\par
\noindent
or \qquad $l(x) \rightarrow  e^{\Lambda} l(x),\quad l = e, \mu, \tau$ ,
i.e. lepton numbers are not conserved separately (i.e. neutrino is mixed)
but there appears a lepton number $l$ related with the common gauge
transformation which is conserved. \\

\par
{\bf 2.4 Schemes of Neutrino Oscillations}\\
\par
Let us consider different schemes of neutrino oscillations.\\
\par
{\bf 2.4.a. Neutrino-Antineutrino Oscillations}\\
\par
The suggestion that, by analogy with $K^o, \bar K^o$ oscillations, there
could be $ \nu, \bar \nu$  oscillations, was considered by B.  Pontecorvo
in work [1].
\par
In this case the mass lagrangian of neutrinos has the following form:
$$
\begin{array}{c} { \cal L}^{'}_{M} =
 - \frac{1}{2} (\bar \nu_e,  \nu_e)
\left(\begin{array}{cc} m_{\nu_e  \nu_e} & m_{\bar \nu_e \nu_e} \\
m_{\nu_e \bar \nu_e} & m_{\bar \nu_e \bar \nu_e} \end{array} \right)
\left(\begin{array}{c} \nu_e \\ \bar \nu_e \end{array} \right) \end{array} .
\eqno(30)
$$
Diagonalizing this mass matrix by standard methods one obtains the following
expression:
$$
\begin{array}{c}{\cal L}^{'}_{M} =
 -\frac{1}{2} (\bar \nu_1, \bar \nu_2)
\left(\begin{array}{cc} m_{\nu_1} & 0 \\
0 & m_{\bar \nu_2} \end{array} \right)
\left(\begin{array}{c} \nu_1 \\ \nu_2 \end{array} \right) \end{array} ,
\eqno(31)
$$
where
$$
\nu_1 = cos \theta \nu_e - sin \theta \bar \nu_e  ,
$$
$$
\nu_2 = sin \theta \nu_e + cos \theta \bar \nu_e   .
$$
These neutrino oscillations are described by expressions (14)-(19) with the
following substitution of $ \nu_\mu \to \bar \nu_e$.
\par
It is necessary to remark that if these neutrinos are Dirac ones, then
the probability to observe $\bar \nu_e$ is much smaller than the probability
to observe $\nu_e$ (such neutrinos can be named the "sterile" neutrinos
(see ref. [3]). It is clear that in this case the lepton
numbers are not conserved, i.e. gauge invariance is violated since the particle
transforms into antiparticle in contrast to the $\nu_e \to \nu_\mu$ transitions
where only aromatic numbers are violated.\\

\par
{\bf 2.4.b. Oscillations of aromatic neutrinos}\\
\par
In the work [2] Maki et al. supposed that there  could exist
transitions between aromatic neutrinos $\nu_e, \nu_\mu$.
Afterwards $\nu_\tau$ was found and then $\nu_e, \nu_\mu, \nu_\tau$
transitions could be possible.
The author of this work has developed this direction (see [30]). It is necessary
to remark that only this scheme of oscillations is realistic
for neutrino oscillations (see also this work).
The expressions which described neutrino oscillations in this case are given above
in expressions (14)-(19).\\

\par
{\bf 2.4.c. Majorana neutrino oscillations}\\
\par
Before discussion of neutrino oscillations in this scheme we give
definitions of Majorana neutrinos (more common consideration, in a
formal form, of this question is given in [6, 24]). Majorana
fermion in Dirac representation has the following form [24, 31]:
\par
$$
\chi^M = \frac{1}{2} [\Psi(x) + \eta_C\Psi^{C}(x)] ,
\eqno(32)
$$
$$
\Psi^C(x) \rightarrow \eta_C C \bar \Psi^T(x) ,
$$
\par
\noindent
where $\eta_{C}$ is a phase, $C$  is a charge conjunction, $T$
is a transposition.
\par
From Exp. (32) we see that Majorana fermion $\chi^M$ has two spin projections
$\pm \frac{1}{2}$ and then the Majorana spinor can be rewritten in the
following form:
\par
$$
\chi^M (x) = \left(\begin{array}{c} \chi_{+\frac{1}{2}}(x)\\
\chi_{-\frac{1}{2}}(x) \end{array} \right) .
\eqno(33)
$$
The mass Lagrangian of Majorana neutrinos in the case of two neutrinos
$\chi_e, \chi_\mu$ ($-\frac{1}{2}$ components of Majorana neutrinos, and
$\bar \chi_{...}$ is the same Majorana fermion with the opposite spin projection)
in the common case has the following form:
$$
\begin{array}{c} { \cal L}^{'}_{M} =
 - \frac{1}{2}(\bar \chi_e, \bar \chi_\mu)
\left(\begin{array}{cc} m_{\chi_e} & m_{\chi_e \chi_\mu} \\
m_{\chi_\mu \chi_e} & m_{\chi_\mu} \end{array} \right)
\left(\begin{array}{c} \chi_e \\ \chi_\mu \end{array} \right) \end{array} .
\eqno(34)
$$
Diagonalizing this mass matrix by standard methods one obtains the following
expression:
$$
\begin{array}{c}  {\cal L}^{'}_{M} =
 - \frac{1}{2}(\bar \nu_1, \bar \nu_2)
\left(\begin{array}{cc} m_{\nu_1} & 0 \\
0 & m_{\nu_2} \end{array} \right)
\left(\begin{array}{c} \nu_1 \\ \nu_2 \end{array} \right) \end{array} ,
\eqno(35)
$$
where
$$
\nu_1 = cos \theta \chi_e - sin \theta \chi_\mu ,
$$
$$
\nu_2 = sin \theta \chi_e + cos \theta \chi_\mu   .
$$
These neutrino oscillations are described by expressions (14)-(19) with the
following substitution of $ \nu_{e \mu}  \to \chi^M_{e \mu}$ .
\par
The standard theory of weak interactions is constructed on the base of local
gauge invariance of Dirac fermions. In this case Dirac fermions have the
following lepton numbers $l_{l,}$ which are conserved (however,
see Sect. 2.1),
\par
$$
l_{l}, l = e ,\mu , \tau,
\eqno(36)
$$
\noindent
and Dirac antiparticles have lepton numbers with the opposite sign
\par
$$
\bar{l} = - l_{l}.
\eqno(37)
$$
\par
Gauge transformation of Majorana fermions can be written in the form:
$$
{\chi'}_{+\frac{1}{2}}(x) = exp(-i\beta) \chi_{+\frac{1}{2}}(x) ,
$$
$$
{\chi'}_{-\frac{1}{2}}(x) = exp(+i\beta) \chi_{-\frac{1}{2}}(x)  .
\eqno(38)
$$
Then lepton numbers of Majorana fermions are
\par
$$
l^{M} =\sum_{i} l^{M}_{i} (+1/2) = -\sum_{i} l^{M}_{i}(-1/2) ,
$$
\noindent
i. e., antiparticle of Majorana fermion is the same fermion with the opposite
spin projection.
\par
Now we come to discussion of the problem of the place of Majorana fermion in
the standard theory of weak interactions [32].
\par
To construct the standard theory of weak interactions [33]
Dirac fermions are used. The absence of contradiction of this theory with the
experimental data confirms that all fermions are Dirac particles.
And in this theory there are numbers which can be
connected with conserving currents. As it is stressed above these numbers
are violated (Sect. 2.1).
\par
Now, if we want  to include the Majorana fermions
into the standard theory we
must take into account that, in the common case, the gauge charges of the Dirac
and Majorana fermions are different (especially it is well seen in the example
of Dirac fermion having an electrical charge since it cannot have a Majorana
charge (it is worth to remind that in the weak currents the fermions are
included in the couples form)). In this case we cannot just include Majorana
fermions in the standard  theory of weak interactions by gauge invariance
manner. Then in the standard theory the Majorana fermions cannot
appear. \\

\par
{\bf 2.4.d. Neutrino Oscillations in the case of Dirac-Majorana
\par
mixing type}\\
\par
We do not discuss this mechanism due to the reason mentioned above.
Consideration of this mechanism can be found in [24].\\
\par
{\bf 2.4.e. Neutrino Oscillation Enhancement in Matter}\\

\par
At present there exist two mechanisms of neutrino oscillation
enhancement in matter. A short consideration of these mechanisms
is given below.\\
\newpage
\par
{\bf 2.4.e.1. Resonant Mechanism of Neutrino Oscillation
\par
Enhancement in Matter}\\

\par
In  strong and electromagnetic interactions the left-handed and
right-handed components of spinors participate in a symmetric manner. In
contrast to these interactions only the left-handed
components of spinors participate in the weak interactions as it is
mentioned above. This is a distinctive feature of the weak interactions.
\par
In the ultrarelativistic limit, the evolution equation for
the neutrino wave function $\nu_{\Phi} $  in
matter has the form [8], [9]
\par
$$
i \frac{d\nu_{Ph}}{dt} = ( p\hat I +
\frac{ {\hat M}^2}{2p} + \hat W ) \nu_{Ph} ,
\eqno(39)
$$
\par
\noindent
where
$p, \hat M^{2}, \hat W $ are, respectively, the momentum,
the (nondiagonal) square mass matrix in
vacuum, and  the  matrix,  taking  into account
neutrino interactions in matter,
$$
\nu_{Ph} = \left (\begin{array}{c} \nu_{e}\\
\nu_{\mu} \end{array} \right) ,
\qquad
\hat I = \left( \begin{array}{cc} 1&0\\0&1 \end{array} \right) ,
$$
$$
\hat M^{2} = \left( \begin{array}{cc} m^{2}_{\nu_{e}\nu_{e}}&
m^{2}_{\nu_{e} \nu_{\mu}}\\ m^{2}_{\nu_{\mu}\nu_{e}}&
m^{2}_{\nu_{\mu} \nu_{\mu}} \end{array} \right).
$$
\par
If we suppose that neutrinos in matter behave analogously to the
photon in matter and the neutrino refraction indices are defined by the
expression
$$
n_{i} = 1 + \frac{2 \pi N}{p^{2}} f_{i}(0) = 1 + 2 \frac{\pi W_i}{p} ,
\eqno(40)
$$
(where $i$ is a type of neutrinos
$(e, \mu, \tau)$, $N$ is density of matter, $f_{i}(0)$ is a real part of
the forward scattering amplitude), then $W$ characterizes polarization of matter by
neutrinos (i.e. it is the energy of matter polarization).
\par
The electron neutrino ($\nu_{e}$)  in
matter interacts via $W^{\pm}, Z^{0}$ bosons and $\nu_{\mu}, \nu_{\tau}$
interact only via $Z^{0}$ boson. These differences in interactions lead to
the following differences in the refraction coefficients of $\nu_{e}$ and
$\nu_{\mu}, \nu_{\tau}$
$$
\Delta n = \frac{2 \pi N}{p^{2}} \Delta f(0) ,
\eqno(41)
$$
$$
\Delta f(0) = - \sqrt{2} \frac{G_F}{2 \pi} ,
$$
where $G_F$ is the Fermi constant.
\par
Therefore the velocities (or effective masses) of
$\nu_{e}$ and $\nu_{\mu}, \nu_{\tau}$
in matter are different. And at the suitable density of matter
this difference can lead to a resonance enhancement of neutrino
oscillations in "matter" [8, 34].
\par
As we can see from the form of Eq. (39), this equation holds
the left-right symmetric neutrinos wave function
$\Psi(x) = \Psi_L(x) + \Psi_R(x)$.
This equation contains the term $W$, which arises from the weak
interaction (contribution of $W$ boson) and which contains only a left-side
interaction of the neutrinos, and is substituted in the left-right
symmetric equation (39) without indication of its left-side origin.
Then we see that equation (39) is an equation that includes term $W$
which arises not from the weak interaction but from a hypothetical
left-right symmetric interaction (see  also works [10, 30, 35]).
Therefore this equation is not the one for neutrinos passing
through real matter. The problem of neutrinos passing through real matter has
been considered in [10, 30, 35, 36].
\par
In three different approaches: by using mass Lagrangian [35, 30], by using
the Dirac equation [35, 30], and using the operator formalism [36],
the author of this work has discussed the problem of the mass generation in
the standard weak interactions and has come to a conclusion that
the standard weak interaction cannot generate
masses of fermions since the right-handed components of fermions do not
participate in these interactions.
Also it is shown [37] that the equation for Green function of the
weak-interacting fermions (neutrinos) in the matter coincides with the equation
for Green function of fermions in vacuum and the law of conservation of
the energy and the momentum of neutrino in matter will be fulfilled [36]
only if the energy $W$ of polarization of matter by the neutrino or the
corresponding term in Wolfenstein equation, is zero (it means that
neutrinos cannot generate permanent polarization of matter).  These results
lead to the conclusion:  resonance enhancement of neutrino oscillations in
matter does not exist.
\par
The simplest method to prove the absence of the resonance enhancement of neutrino
oscillations in matter is:
\par
   If we put an electrical (or strong) charged particle in  matter,
there arises polarization of matter. Since the field around the particle
is spherically symmetrical, the polarization must also be spherically
symmetrical. Then the particle will be left at rest and the law of
energy and momentum conservation is fulfilled.
\par
If we put a weakly interacting particle (a neutrino) in matter then,
since the field around the particle has a left-right asymmetry (weak
interactions are left interactions with respect to the spin direction),
polarization of matter must be nonsymmetrical, i.e. on the left side
there arises maximal polarization and on the right there is zero polarization.
Since polarization of the matter is asymmetrical, there arises asymmetrical
interaction of the particle (the neutrino) with matter and the
particle cannot be at rest and will be accelerated. Then the law
of energy momentum conservation will be violated. The only
way to fulfil the law of energy and momentum conservation
is to demand that polarization of matter be absent in the weak interactions.
The same situation will take place in vacuum.
\par
It is interesting to remark that in the gravitational interaction
the polarization does not exist either [38].\\

{\bf 2.4.e.2. Enhanced Oscillation of Neutrinos of Different
\par
Masses in Matter}\\

\par
The oscillation  probability  is  estimated  for  neutrinos  of
different  masses  in their  passing   through  matter of  different
thickness, including the Sun [13, 27].
\par
1) So, if neutrinos of different  types  have  equal  masses,  real
oscillations are  possible  for  different  types  of  neutrinos  by
analogy with $K^o , \bar K^o$  oscillation;
\par
2) if neutrino masses  are  different  for  different  neutrino
types, only virtual neutrino oscillations are  possible  while  real
oscillations require participation of neutrinos in interactions  for
their transition to the  respective  mass  shells  by  analogy  with
transition of a $\gamma$-quantum to the $\rho$ -meson in  the  vector
dominance model.
\par
 We shall estimate  the  probability
for neutrinos to change from one type $\nu_l$ to another
$\nu_{l'} (m_{\nu_{l}} \neq  m_{\nu_{l'}})$
in passing through matter.  Neutrino  transition  to the mass  shell
will occur via the weak neutrino-matter  interaction  (by  analogy  with
the $\gamma-\rho^o$  transition or $K^o_1 , \bar K^o_2$  oscillation).
We shall assume  that difference in mass of $\nu _{l} , \nu _{l'}$
neutrinos is small enough to  consider $\nu _{l'}$
the probability of transition to the mass shell proportional  to
 the total elastic cross section $\sigma^{el} (p)$   for the  weak
interaction (for simplicity we shall deal with the oscillation of  two
types of neutrinos).  Then the length of the elastic interaction of  the
neutrino  in  the  matter of density  charge $Z$, atomic number $A$
and  momentum  $p$  will  be defined as
$$
\Lambda_{0} \sim {1\over \sigma^{el}(p) \rho  (z/A)} .
$$
If the neutrino mass difference is fairly large,  it  can  be  taken
into account by the methods of the vector dominance  model  [39].  As
pointed out above, we shall assume  that  this  difference  is  very
small and employ above formula.
\par
The real part of forward scattering amplitude  $Re f_i (p, 0)$  is
responsible  for  elastic  neutrino  scattering  in matter (it is supposed
that at low energies the coherent process takes place).  It  is related
to the exponential phase term $exp(-p \Delta_i r)$ (as factor to
momentum) in the wave function of particle $\Psi(r,...)$ and has the following
form:
$$
p \Delta_{i} \simeq  {2\pi  N_{e} f_{i} (p,0)\over p} ,\qquad
i = \nu _{e}, \nu _{\mu }, \nu _{\tau} .
\eqno(42)
$$
Keeping in mind that [33]
$$
f_{i} (p,0) \simeq  \sqrt{2} G_{F} p\left({M^{2}_{W}\over M^{2}_{i}}\right)  ,
\eqno(43)
$$
\par
if $i = \nu _{e},\qquad M^{2}_{i} = M^{2}_{W}$ ,\\
\par
if $i = \nu_{\mu}, \nu_{\tau},\qquad M^{2}_{i} = M^{2}_{{Z}^0}$  , \\
\par
\noindent
we obtain
$$
p(\Delta_{i}) \simeq \sqrt{2} G_{F} N_{e}
\left({M^{2}_{W}\over M^{2}_{i}}\right).
$$
The phase of the elastic scattering amplitude changes by $2\pi $
over the length
\par
$$
\Lambda_{i 0} \simeq  {2\pi \over \sqrt{2} G_{F} \rho(z/A)
\left({M^{2}_{W}\over M^{2}_{i}}\right)} = 2\pi  L_{i 0}  \sim \Lambda_0.
\eqno(44)
$$
For simplification further we will suppose that
$M^2_e \simeq M^2_\mu \simeq M^2_\tau$ and then $\Lambda_i = \Lambda$.
(Absorption  or  the  imaginary  part  of  the  forward  scattering
amplitude can be ignored for low-energy neutrinos.)
\par
Knowing  that   the  length  of  elastic  neutrino-matter
interaction is $\Lambda_0$, we must estimate the oscillation
probability for the  neutrino  passing  through  the  matter  of
thickness $L$. The probability of the elastic $\nu _{l}$  interaction in
matter of thickness $L$ is
$$
 P( L ) = 1- exp ( -2 \pi L/ \Lambda_0) .
\eqno(45)
$$
Then, using formulae (44), (45), we can find the neutrino oscillation
probability $\rho_{\nu_{l} \nu_{l'} } (L)$  at  different  thickness $L$.
Averaging  the expression for neutrino oscillation probability [13]
over $R$
$$
P_{\nu_{l}\nu_{l'}}(R) = {1\over 2} sin^{2} 2\theta_{\nu_{l}\nu_{l'}}
(1 - cos  2\pi  {R\over L_0}) ,
\eqno(46)
$$
where  $L_{0} = {4\pi p\over \Delta m^{2}}$  , \\
\par
\noindent
then we obtain
\par
$$
\bar P_{\nu_{l}\nu_{l'}}(R) = {1\over 2} \sin^{2} 2\theta_{\nu_{l}\nu_{l'}} .
$$
Then the oscillation probability $\rho_{\nu_{l} \nu_{l'} } (L)$
or the mixing angle  $\beta $  at $\Lambda_0 \ge L_0$ will be defined
by the expressions (for simplicity it is supposed  that
$\Lambda_0 = \Lambda_e = \Lambda_\mu = \Lambda_\tau$):
\par
a) for $L$ comparable with $\Lambda_0$,
\par
$$ \rho_{\nu_{l}\nu_{l'}}(L) = {1\over 2} \sin^{2} 2\beta \simeq
\bar P_{\nu_{l}\nu_{l'}} = {1\over 2} \sin ^{2}
2\theta_{\nu_{l}\nu_{l'}} ,
\eqno(47)
$$
where $\beta \simeq  \theta_{\nu_{l}\nu_{l'}}$ ;
\par
b) for very large $L, {L\over \Lambda_{0}} >
{1\over \sin^{2} 2\theta_{\nu_{l}\nu_{l'}}} \gg  1$,
$$
\rho_{\nu_{l}\nu_{l'}}(L) = {1\over 2} \sin^{2} 2\beta  \simeq  {1\over 2} ,
\eqno(48)
$$
and $\beta  \simeq  {\pi \over 4}$ ;\\
\par
c) for intermediate L,
$$
\rho_{\nu_{l}\nu_{l'}}(L) =
{1\over 2} \sin ^{2} 2\theta_{\nu_{l}\nu_{l'}} \le  \rho_{\nu_{l}\nu_{l'}}(L)
 \le  {1\over 2}   ,
\eqno(49)
$$
and \qquad $\theta_{\nu_{l}\nu_{l'}} \le  \beta  \le  {\pi \over 4}$ .
\par
If $L_{0} \ge  \Lambda _{0}$ the expressions like (47)-(49) are also
true, but $\Lambda_{0}$ should  be  replaced by $L_{0}$ and the
thickness  of  matter   will  be determined in units of $L_{0}$.
Also, since  the  oscillation  length  $L_{0}$ increases with
the  neutrino  momentum  (see (46)),  the  number  of
oscillation lengths $n = L/L_{0}$ fitting  in  the  given  thickness
$L$ decreases with increasing neutrino  momentum  as, accordingly, the
neutrino oscillation probability $\rho_{\nu_{l}\nu_{l'}}(L)$ does.
\par
Let us consider the neutrino oscillation probability
for intermediate interaction numbers $n$. The distribution probability
of  $n$-fold  elastic  neutrino  interaction  for  thickness $L$  with
the mean  value $\bar n = L/ \Lambda_0$ at  not very large $\bar n$  is
determined by the  Poisson distribution
$$
f(n,\bar n) = {(\bar n)^{n}\over n!} e^{-\bar n} .
\eqno(50)
$$
At large $\bar n$ it changes to the Gaussian distribution:
$$
f(n, \bar n, \bar n) = {1\over \sqrt{2\pi \bar n}}
e^{{-(n - \bar n)^2 \over 2\bar n}}   .
\eqno(51)
$$
The probability of neutrino conversion from $\nu_{l}$ to $\nu_{l}$ and
$\nu_{l'}$ $n$-fold  elastic  interaction  is determined  by  recursion
relations (where $\theta  \equiv  \theta_{\nu_{l} \nu_{l'} }$)  given in
works [13, 27].
\par
Here we give the expression for probability for neutrino conversion
at  $\sin^{2} 2\theta  \ll  1$ for two types of neutrinos ($\nu_e, \nu_\mu$)
$$
\rho (\nu_{e} \rightarrow  \nu_{e}) = 1 -\bar n {1\over 2} \sin^{2} 2\theta  ,
\eqno(52)
$$
$$
\rho (\nu_{e} \rightarrow  \nu_{\mu }) = \bar n {1\over 2} \sin^{2} 2\theta  .
$$
Then enhancement of neutrino oscillation in matter
will take place, i.e. $\nu_{e}$ neutrinos will transit
in $\nu_{\mu }, \nu_{\tau}$ neutrinos, but it is necessary to take into
account that mean numbers of interaction lengths $L^o_\mu, L^o_\tau$
of  $\nu_{\mu }, \nu_{\tau}$ will be  $\delta $ times
less and then, correspondingly,  $\bar n$  in (52) will be
changed for $\bar n_{\mu }, \bar n_{\tau}$.
$$
\delta = \bar n_{e}/ \bar n_{\mu }  = \bar n_{e}/ \bar n_{\tau} \simeq  2.49 .
\eqno(53)
$$
\par
The mean number  of
elastic interactions of electron neutrinos produced in the Sun is
$$
\Lambda_{Sun} \simeq 1.7 \cdot 10^7 m, \qquad \bar n^{Sun}_e \simeq 40,
\qquad \bar n^{Sun}_\mu \simeq 16,  \bar n^{Sun}_\tau  \simeq 16.
$$
\par
It is necessary to mention that the considered mechanisms of enhancement of
neutrino oscillation in matter lead only to changing the mixing angles and for
their realizations the vacuum mixing angle of neutrino oscillations must differ
from zero.\\

\par
{\bf 2.4.f. Neutrino Oscillations in Supersymmetrical Models}\\

\par
Neutrino oscillation in supersymmetrical models is considered in works (and
see references there)
[40-42]. Here we do not fulfil detailed considerations of these schemes
but want to remark that in these schemes side by side the neutrino
oscillations the superpartner of the fermions and bosons
must be observed.\\

\section{ Experimental Data}

\par
{\bf 3.a. Neutrino Experimental Data from SNO (Canada)}\\
\par
The SNO detector [18], containing 1000 tons of heavy water ($D_2 O$), is
placed in a shaft in Sudbury  at depth 6010 m water equivalent (Sudbury
Neutrino Observatory).
\par
The neutrinos are detected in the following reactions:
\par
\begin{center}
\begin{tabular}{lc}
1. $\nu _{x} + e^{-} \rightarrow \nu _{x} + e^{-}$, &$E_{thre} \simeq$ 6 MeV (ES), \\
2. $\nu _{e} + d \rightarrow p + p + e^{-}$  ,   & $E_{thre} \simeq$ 1.45 MeV (CC),\\
3. $\nu_{x} + d \rightarrow p + n + \nu_{x}$ , & $E_{thre} \simeq$ 2.23 MeV (NC), \\
$x = e, \mu , \tau$  .                           &                          \\
\end{tabular}
\end{center}

\par
Reaction 1 goes through charged and neutral currents, if
$x = e$, and neutral if $x = \mu , \tau$; reaction 2 goes through charged
current, and reaction 3 through neutral current. Using any couple of the
reactions we can find the primary flux of the Sun neutrinos. Sudbury reported
first results in [20, 43]. These results are obtained for the Sun neutrinos with
threshold $E_{eff} \ge 6.75 MeV$.
\par
Figure 1 shows the distribution of $cos\theta$ (a), and kinetic
energy spectrum with a statistical error (b) with the $^{8}B$ spectrum [44]
scaled to the data. The ratio of the data to the prediction [45] is shown
in (c). The bands represent the $1 \sigma$ uncertainties derived from the
most significant energy-dependent systematic errors. There is no evidence
for a deviation of the spectral shape from the predicted shape on the
non-oscillation hypothesis.
\par
Normalized to the integrated rates above the energy $E_{eff} = 6.75$ MeV, the
flux of neutrinos is (from reactions 2 and 1):
$$
\phi^{CC}_{SNO} (\nu_e) = 1.75\pm0.07(stat.)+0.12(-0.11)(syst.)
\eqno(54)
$$
\par
$\pm0.05(theor.) \times 10^6 cm^{-2}s^{-1}$ ,
$$
\phi^{ES}_{SNO} (\nu_x) = 2.39\pm0.34(stat.)+0.16(-0.14)(sys.)
\times 10^6 cm^{-2}s^{-1} ,
\eqno(55)
$$
where the theoretical uncertainty is the CC cross section uncertainty.
The neutrinos flux (55) measured on SNO is consistent with the same flux
measured on Super-Kamiokande (58).

\begin{center}
\mbox{\epsffile{sno1.eps}} \vspace{2mm} \noindent
\end{center}
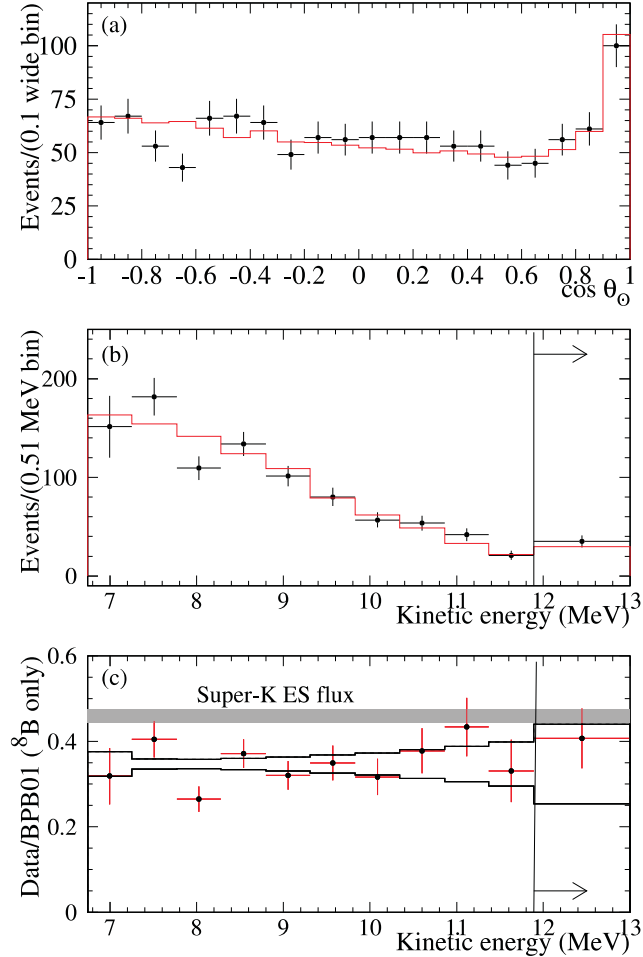
 {\sf Figure~1:}~Distributions of (a) $cos \theta_{sun}$, and (b) extracted
kinetic energy spectrum for CC events with $R \le 5.50$ m and
$T_{eff} \ge$ 6.75 MeV. The Monte Carlo simulations for an undistorted
$^{8}B$ spectrum are shown as histograms. The ratio of the data to the expected
kinetic energy distribution with correlated systematic errors is shown in (c).
The uncertainties in the $^{8} B$ spectrum have not been included



\par
The difference between $\nu$ flux deduced from the ES rate and that deduced
from the CC rate is
$$
\phi_{SNO} = 0.64\pm0.40 \times 10^6 cm^{-2}s^{-1}  .
$$
It is the $\nu_\mu , \nu_\tau$ flux measured through NC.
\par
The best fit to the $\phi_{SNO} (\nu_{\mu \tau})$ flux is:
$$
\phi_{SNO} (\nu_{\mu \tau}) = 3.69\pm1.13 \times 10^6 cm^{-2}s^{-1} .
\eqno(56)
$$
\par
The ratio of the SNO CC flux to the solar model [44 ] is
$$
\frac{\phi_{SNO}^{CC}}{\phi_{BPB00}} = 0.347\pm0.029 .
$$

\par
The total flux of active $^{8}B$ neutrinos is determined to be:
$$
\phi_{SNO} (\nu_x) = 5.44\pm0.99 \times 10^6 cm^{-2}s^{-1} .
\eqno(57)
$$
This result is in a good agreement with prediction of the standard solar
models [45, 46].
\par
The SNO results are the first direct indication of the non-electron flavor
components in the solar neutrino fluxes, and it is, practically [45, 46],
the total flux of $^{8}B$ neutrinos generated by the Sun.\\

\par
{\bf 3.b. Neutrino Experimental Data from Super-Kamiokande
\par
(Japan)}\\
\par
The Super-Kamiokande detector [15, 16] is a cylindrically-sharped water
Cherenkov detector with 50000 ton of ultra-pure water. It is located about
1000m (2700 m.w.e.) underground in the Kamioka mine. Super-Kamiokande is a
multipurpose experiment, and solar and atmospheric neutrino physics is one of its
main topics.

\par
i). The Sun neutrino fluxes measured in Super-Kamiokande detector [47] through
the electron scattering reaction
1. $\nu _{x} + e^{-} \rightarrow \nu _{x} + e^{-}$ $…_{thre} \simeq$ 5 MeV
are as follows:
$$
\phi^{ES}_{SK} (\nu_e) = 2.32\pm0.03(stat.)+0.08(-0.07)(syst.)
\times 10^6 cm^{-2}s^{-1} .
\eqno(58)
$$
These fluxes are in a good consistence with the Sun neutrino fluxes measured in SNO.

\begin{center}
\mbox{\epsffile{sk1.eps}} \vspace{2mm} \noindent
\end{center}
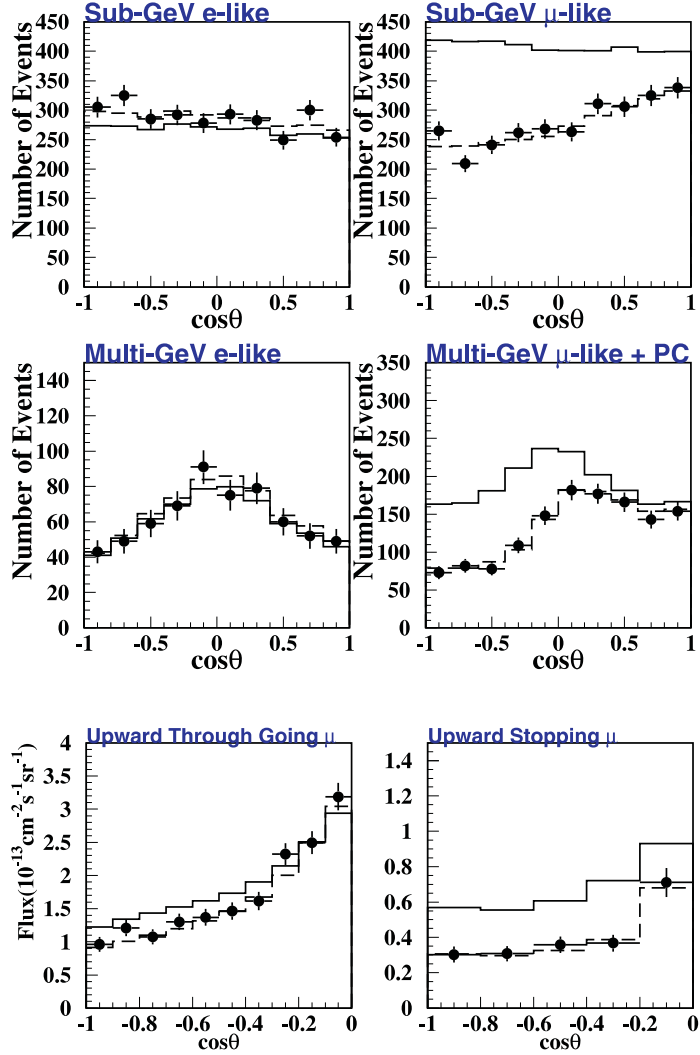
 {\sf Figure~2:}~Zenith angle distribution of Super-Kamiokande 1289
days FC, PC and UPMU samples. Dots, solid and dashed lines correspond to
data, MC with no oscillation and MC with best fit oscillation parameters,
respectively

\vspace{0.5cm}
\par
The day-night asymmetry $A$ is
$$
A = \frac{(\Phi_n - \Phi_d)}{((\Phi_n + \Phi_d)/2)} = 0.033 \pm 0.022(stat.)
+0.013(-0.012)(syst.) .
$$
This is $1.3 \sigma$ from the zero asymmetry.
\par
ii). Atmospheric neutrinos are produced by the interactions of the primary
cosmic rays on nuclei of the Earth atmosphere. The atmospheric neutrinos at
a few GeV have ratio 2 ${(\nu_\mu+ \bar \nu_\mu)/(\nu_e + \bar \nu_e})$.
The events observed in Super-Kamiokande are categorized into
four types: (1) Fully Contained (FC) events, which have their vertex in
the detector and all visible particles contained in the detector.
(2) Partially Contained (PC) events, which have their vertex in the detector
and at least one visible particle exits from the detector. (3) Upward
through-going muons which are produced by the $\nu_\mu$ charged current
interaction in the rock surrounding the detector and go through the detector.
(4) Upward stopping muons which are produced by the $\nu_\mu$ charged current
interaction in the rock surrounding the detector but stop in the detector.
The primary neutrino ($\nu_e$-like and $\nu_\mu$-like) energy are divided in
two regions: (1) $E_\nu \le 1.33$ GeV sub-GeV, (2) $E_\nu> 1.33$GeV multi GeV.
\par
Figure 2 gives the zenith angle distribution of Super-Kamiokande
1289 days samples [48]. Dots, solid and dashed lines correspond to
the data, MC with no oscillation and MC best oscillation
parameters [49], respectively ($\Delta m = 2.5 \times
(10)^{-3}eV^2$, $sin^2 2 \theta = 1.00$) . These data are well
explained by $\nu_\mu \to \nu_\tau$ 2-flavor oscillations and are
consistent with $\nu_\tau$ appearance roughly at the two-sigma
level.

\section{Conclusions from Comparison of the Experimental Data with Theoretical
Scheme Predictions on Neutrino Oscillations}

\par
1. In the Super-Kamiokande experiment on atmospheric neutrinos
the deficit of muonic neutrinos is detected. The analysis shows that
they can transit only
in $\nu_\tau$ neutrinos. The $\nu_\mu \to \nu_e$ transition in this
experiment is not observed. From this fact we can conclude (taking
into account SNO results) that the length of $\nu_\mu \to \nu_\tau$ transitions
is of the order of the Earth diameter, and the angle $\theta$ of
$\nu_\mu \to \nu_\tau$ transitions is near to the maximal mixing angle
$\theta \cong \pi / 4$. Then the length of $\nu_\mu \to \nu_e$ transitions
is much more than the Earth diameter. The SNO experimental data also confirm,
through neutral current registration, $\nu_\mu \to \nu_\tau$ transitions with
the same mixing angle.

\par
2. From the SNO experimental data on straight registration neutrinos
(by neutral and charge currents in case $\nu_e$ and neutral current in
the case $\nu_\mu, \nu_\tau$) we can come to the following conclusion:
the primary $\nu_e$ neutrinos transit in equally proportions in
$\mu_e, \nu_\mu, \nu_\tau$ neutrinos, i. e., mixing angles
$\theta_{(...)}$ of $\nu_e, \nu_\mu, \nu_\tau$ are equal to the maximal
angles of mixing. The length of $\nu_e \to \nu_\mu, \mu_\tau$ oscillations
is less than the distance to the Sun.
\par
Come to comparison of these results with the predictions in the
above-considered theoretical schemes on neutrino oscillations.\\

\par
{\bf 4.a. Neutrino-Antineutrino Oscillations}\\

\par
In the existing experimental results the neutrinos disappearance
has not been detected (see above), i. e., this mechanism is not confirmed.\\

\par
{\bf 4.b. Aromatic Neutrino Oscillations}\\
\par
This scheme was confirmed by experiments.
\par
Pontecorvo-Gribov type oscillations for aromatic neutrinos maximal mixing
angle can be realized only at neutrino masses equality
$m_{\nu_e} = m_{\nu_\mu} = m_{\nu_\tau}$. It is hardly probable that
the neutrino masses are equal.
The length of $\mu_\mu \to \nu_\tau$ oscillations nearly is equal to the
Earth diameter, and the length of $\nu_e \to \nu_\tau$ oscillations is more much
than the Earth diameter. Then more probable is the type of oscillations
suggested by the author [36], $\theta_{(...)} = \pi/4$ (see Sect. 2 and below)
and the transition between oscillating neutrinos is virtual.
Here neutrino oscillations can take place in the charge mixings scheme [50].
It is supposed that the neutrinos are mixed via weak interactions and
therefore if we consider charge mixings of two neutrinos-$a, b$, then the
mixing angles must be
$$
sin \theta \cong \frac{g_w(a)}{\sqrt{g^2_w(a) + g^2_w (b)}} \cong
\frac{1}{\sqrt{2}} ,
$$
since $g_w(a) \cong g_w(b)$, where $g_w(a), g_w(b)$ are weak couple constants
of $a, b$ neutrinos. \\

\par
{\bf 4.c. Majorana Neutrino Oscillations}  \\
\par
From the above-considered discussion (see Sect. 2.4.c) we can come to a conclusion
that the Dirac and Majorana gauge charges are different and therefore we
cannot put Majorana fermions in the Dirac theory.
Then it is obvious that this scheme of neutrino oscillations
cannot be realized.\\

\par
{\bf 4.d. Neutrino Oscillations in the Scheme of Majorana-
\par
Dirac Mixing Type}  \\
\par
We do not discuss this scheme for the reason given above (see Sect. 2.4.c).
It is clear that this scheme cannot be realized in the experiment either.\\

\par
{\bf 4.e. Mechanisms of Neutrino Oscillations Enhancement
\par
in Matter}\\
\par
{\bf 4.e.1 Mechanism of Resonance Enhancement of Neutrino
\par
Oscillations in Matter}\\

\par
The experimental data on energy spectrum and day-night effect obtained in
Super-Kamiokande (energy spectrum of neutrinos is not distorted, day-night
effect is within the experimental mistakes) and the results obtained in SNO  have
not confirmed this effect. Besides, this effect can be realized only at the
violation of the law of the energy-momentum conservation (see Section 2.4.e.1. in
this work and Ref. [26]).\\

\par
{\bf 4.e.2. Mechanism of Accumulation of the Neutrino
\par
Different Masses in Matter} \\
\par
This mechanism effectively works only at small mixing angles. Since the
mixing angles discovered in SNO and Super-Kamiokande are maximal, then we can
neglect the contribution of this mechanism to the neutrino oscillations.\\

\par
{\bf 4.f. Neutrino Oscillation in Supersymmetric Models} \\

\par
This type of oscillations can be confirmed only in case of discovery
of the superpartners of fermions and bosons besides the neutrino oscillations.\\

\section{Conclusion}

\par
The theoretical schemes on neutrino oscillations are considered. The
experimental data on neutrino oscillations from Super-Kamiokande (Japan)
and SNO (Canada) are given. The comparison of these data with theoretical schemes
has been done. We have come to a conclusion: The experimental data confirm only the
scheme with transitions (oscillations) between aromatic
$\nu_e, \nu_\mu, \nu_\tau$ neutrinos with maximal mixing angles.
This scheme was suggested by Z. Makki et al., in 1962 [2] and repeated
by B. Pontecorvo in 1967 [3] and subsequently is developed by  Kh. Beshtoev
(see references in this work). Besides, this mechanism of a neutrino
oscillations is the only one which is theoretically substantiated.\\

\newpage
\par
{\bf References}\\

\par
\noindent
1. Pontecorvo B. M., Soviet Journ. JETP, 1957, v. 33, p.549;
\par
JETP, 1958,  v.34, p.247.
\par
\noindent
2. Maki Z. et al., Prog.Theor. Phys., 1962, vol.28, p.870.
\par
\noindent
3. Pontecorvo B. M., Soviet Journ. JETP, 1967, v. 53, p.1717.
\par
\noindent
4. Davis R. et al., Phys. Rev. Letters, 1968, vol.20, p.1205.
\par
\noindent
5. Bahcall J. et al., Phys. Lett.B, 1968, vol.26, p.1;
\par
Bahcall J.,  Bahcall N., Shaviv  G.,  Phys.  Rev.  Lett.  1968,
vol.20,
\par
p.1209;
\par
   S. Turck-Chiere et al., Astrophys.J. 335 (1988), p.415.
\par
\noindent
6. Gribov V., Pontecorvo B.M., Phys. Lett. B, 1969, vol.28, p.493.
\par
\noindent
7. Hirata K.S. et  al., Phys. Rev. Lett., 1989, vol.63,p.16.
\par
\noindent
8. Mikheyev S.P., Smirnov A.Ju., Nuovo Cimento, 1986, vol.9,p.17.
\par
\noindent
9. Wolfenstein L., Phys. Rev.D, 1978, vol.17, p.2369.
\par
\noindent
10. Beshtoev Kh.M., JINR Commun. E2-91-183, Dubna, 1991;
\par
Proceedings of III  Int. Symp. on Weak and Electromag. Int. in
\par
Nucl. (World Scient., Singapoure, P. 781, 1992);
\par
13th European Cosmic Ray Symp. CERN, Geneva, HE-5-13.
\par
\noindent
11. Anselmann P. et al., Phys. Lett. B, 1992,
\par
vol.285, p.376;  1992, vol.285, p.391;
\par
Hampel W. et al., Phys. Lett. B, 1999, v.447, p. 127.
\par
\noindent
12. Abdurashitov  J.N.  et al.,  Phys.  Lett.B,  1994,  vol.328,
\par
p.234; Phys. Rev. Lett., 1999, v.83,  p.4683.
\par
\noindent
13. Beshtoev Kh.M., JINR Commun, E2-93-297, Dubna, 1993;
\par
JINR Commun. E2-94-46; Hadronic Journal, 1995, vol 18, p.165.
\par
\noindent
14. Beshtoev Kh.M., JINR Commun. E2-96-458, Dubna, 1996;
\par
JINR Commun. E2-97-360, Dubna, 1997;
\par
Report at International Conf. "Neutrino98", Japan, 1998.
\par
\noindent
15. Totsuka Y., Proc. Intern. Symp.  on  Underground  Exp.  (ed.K.
\par
Nakamura), Tokyo, 1990, p.129.
\par
\noindent
16. Suzuki Y., Report at Intern. Conf. "Neutrino98", Japan,
\par
June, 1998; Phys. Rev. Lett. 1998, v.81 p.1158;
\par
Fukuda Y. et al., Phys. Rev. Lett., 1999, v.82, p.2430;
\par
1999, v.82, p.1810.
\par
\noindent
17. Kajita T., Report on Intern. Conf. "Neutrino98" Japan,
\par
June, 1998;
\par
Fukuda Y. et al., Phys. Rev. Lett., 1999,
\par
v.82, p.2644.
\par
\noindent
18. Aardsma et al., Phys. Lett. B, 1987, vol.194, p.321.
\par
\noindent
19. Kameda J., Proceedings of ICRC 2001, August 2001, Germany,
\par
Hamburg, p.1057.
\par
\noindent
20. Ahmad Q. R. et al., Internet Pub. nucl-ex/0106015, June 2001.
\par
\noindent
21. Bogolubov N.N., Shirkov D.V., Introd. to the Quantum Field
\par
Theory, M.: Nauka, 1986;
\par
Kane G., Modern Elementary Particle Physics,
\par
Add. W. P.C., 1987.
\par
\noindent
22. Beshtoev Kh.M., JINR Commun. E2-99-81, Dubna, 1999;
\par
hep-ph/99 .
\par
\noindent
23. Beshtoev Kh. M., INR AC USSR Preprint -577, Moscow, 1988.
\par
\noindent
24. Bilenky S.M., Pontecorvo B.M., Phys. Rep., C41(1978)225;
\par
Boehm F., Vogel P., Physics of Massive Neutrinos: Cambridge
\par
Univ. Press, 1987, p.27, p.121;
\par
Bilenky S.M., Petcov S.T., Rev. of Mod.  Phys., 1977, v.59,
\par
p.631.
\par
\noindent
25. Beshtoev Kh.M., JINR Commun. E2-92-318, Dubna, 1992;
\par
JINR Rapid Communications, N3[71]-95.
\par
\noindent
26. Beshtoev Kh.M., Internet Pub. hep-ph/9911513;
\par
 The Hadronic Journal, v.23, 2000, p.477;
\par
Proceedings of 27th Intern. Cosmic Ray Conf., Germany,
\par
Hamburg, 7-15 August 2001, v.3, p. 1186.
\par
\noindent
27. Beshtoev Kh.M., Hadronic Journal, 1995, vol.18, p.165.
\par
\noindent
28. Blatt J.M., Waiscopff V.F., The Theory of Nuclear Reactions,
\par
INR T.R. 42.
\par
\noindent
29. Beshtoev Kh.M., JINR Commun. E2-99-307, Dubna, 1999;
\par
JINR Commun. E2-99-306, Dubna, 1999.
\par
\noindent
30. Beshtoev Kh.M., Phys. of Elem. Part. and Atomic Nucl.
\par
(Particles and Nuclei), 1996, v.27, p.53.
\par
\noindent
31. Rosen S.P., Lectore Notes on Mass Matrices, LASL preprint,
\par
1983.
\par
\noindent
32. Beshtoev Kh.M., JINR Commun. E2-92-195, Dubna, 1992.
\par
\noindent
33. Glashow S.L.- Nucl. Phys., 1961, vol.22, p.579 ;
\par
Weinberg S.- Phys.  Rev. Lett., 1967, vol.19, p.1264 ;
\par
Salam A.- Proc. of the 8th Nobel  Symp.,  edited  by
\par
N. Svarthholm (Almgvist and Wiksell,  Stockholm) 1968, p.367.
\par
\noindent
34. Mikheyev S. P., Smirnov A. Yu., Yad. Fiz. 1986, v.42, p.1441;
\par
Sov. Phys. JETP, 1986, v.91, p.7;
\par
Mikheyev S. P., Smirnov A. Yu, Nuovo Cimento  C 9, 1986, p.17;
\par
Boucher J. et al., Z. Phys. C 32, 1986, p.499.
\par
\noindent
35. Beshtoev Kh. M., JINR Communication E2-93-167, Dubna,
\par
1993; JINR Communication P2-93-44, Dubna, 1993;
\par
\noindent
36. Beshtoev Kh.M., HEP-PH/9912532, 1999;
\par
Hadronic Journal, 1999, v.22, p.235.
\par
\noindent
37. Beshtoev Kh.M., JINR Communication E2-2000-30, Dubna,
\par
2000; Internet Publ. hep-ph/0003274.
\par
\noindent
38. Beshtoev Kh.M., JINR Communication P2-2001-65, Dubna,
\par
2001.
\par
\noindent
39. Sakurai J.J., Currents and Mesons. The Univ. of Chicago
\par
Press, 1967.
\par
\noindent
40. Haug et al., Nucl. Phys. B565, 2000, p.3848.
\par
\noindent
41. Bednyakov V. et al., Nucl. Phys. B442, 1998, p.203.
\par
\noindent
42. Dib C. et al., hep-ph/0011213.
\par
\noindent
43. C. Waltham, Proceedings of ICRC 2001,
August 2001,
\par
Hamburg, Germany, v.4, p.3167.
\par
\noindent
44. Ortz C.E. et al., Phys. Rev. Lett., 2000, v.85, p.2909.
\par
\noindent
45. Bahcall J.N. et al., astro-ph/0010346.
\par
\noindent
46 Turck-Chieze S. et al., Ap. J. Lett., v.555 July 1, 2001.
\par
\noindent
47. Fukuda S. et al., Phys Rev. Lett., 2001, v.86, p.5651.
\par
\noindent
48. Toshito T., hep-ex/0105023;
\par
Kameda J., Proceeding of 27th ICRC, August 2001,
\par
Hamburg, Germany, v.2, p.1057.
\par
\noindent
49. Honda M. et al., Phys Rev. D52, 1995, p.4985; Phys. Rev. D53,
\par
1996, p.1313.
\par
\noindent
50. Beshtoev Kh. M., JINR Communication E2-2000-229, Dubna,
\par
2000.

\end{document}